\documentclass[sigconf, screen=false, review=false, anonymous=false, authordraft=False]{acmart}

\usepackage[yyyymmdd]{datetime}

\usepackage{tabularx}
\usepackage{dcolumn} 
\newcolumntype{d}[1]{D{.}{.}{#1}}
\usepackage{subcaption}
\usepackage{hyperref}
\usepackage{todonotes}
\let\xtodo\todo
\renewcommand{\todo}[1]{\xtodo[inline,color=green!50]{#1}}

\AtBeginDocument{%
  \providecommand\BibTeX{{%
    \normalfont B\kern-0.5em{\scshape i\kern-0.25em b}\kern-0.8em\TeX}}}

\setcopyright{cc}
\acmConference[]{\href{https://humannotebookinteractions.github.io}{1st CHI Workshop on Human-Notebook Interactions}}{May 11, 2024}{Honolulu, HI, USA}

\begin{document}

\title{From Computational to Conversational Notebooks}

\author{Thomas Weber}
\orcid{0000-0002-6894-605X}
\affiliation{%
 \institution{LMU Munich}
 \streetaddress{Frauenlobstr. 7a}
 \city{Munich}
 \country{Germany}
 \postcode{80337}
}
\email{thomas.weber@ifi.lmu.de}

\author{Sven Mayer}
\orcid{0000-0001-5462-8782}
\affiliation{%
 \institution{LMU Munich}
 \streetaddress{Frauenlobstr. 7a}
 \city{Munich}
 \country{Germany}
 \postcode{80337}
}
\email{info@sven-mayer.com}

\renewcommand{\shortauthors}{Weber and Mayer}

\begin{abstract}
Today, we see a drastic increase in LLM-based user interfaces to support users in various tasks. Also, in programming, we witness a productivity boost with features like LLM-supported code completion and conversational agents to generate code. In this work, we look at the future of computational notebooks by enriching them with LLM support. We propose a spectrum of support, from simple inline code completion to executable code that was the output of a conversation. We showcase five concrete examples for potential user interface designs and discuss their benefits and drawbacks. With this, we hope to inspire the future development of LLM-supported computational notebooks.
\end{abstract}



\begin{teaserfigure}
    \centering
    \includegraphics[width=\linewidth]{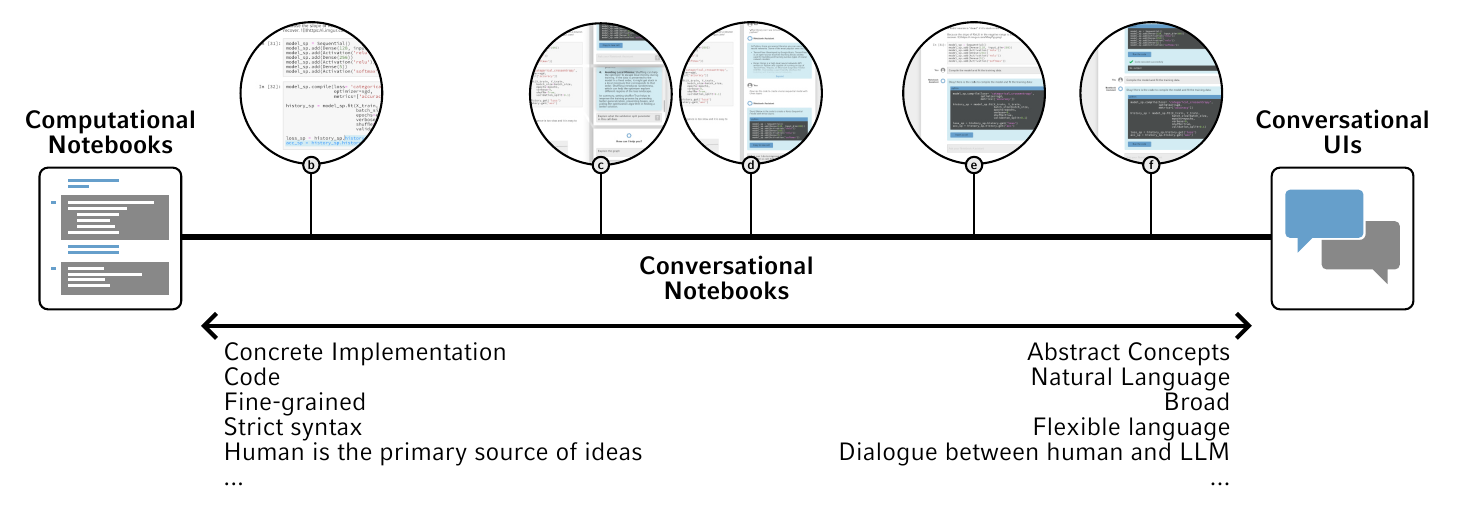}
    \caption{Conversational Notebooks, including the examples described above, fall on a spectrum, depending on how strongly each of the two interface types is represented in the final design. This affects many aspects of the interface and the interaction with it.}
    \label{fig:spectrum}
\end{teaserfigure}

\maketitle

\section{Introduction}
Large Language Models (LLMs) have proven to be a technology that greatly impacts many areas of our lives by enhancing, simplifying, or replacing existing tasks, workflows, or activities. Consequently, many companies now integrate LLMs into their products (e.g., Microsoft Copilot\footnote{\url{https://copilot.microsoft.com}},  Salesforce\footnote{\url{https://www.salesforce.com/news/press-releases/2023/06/12/ai-cloud-news/}}, BloombergGPT\footnote{\url{https://www.bloomberg.com/company/press/bloomberggpt-50-billion-parameter-llm-tuned-finance/}}) to make them and their benefits accessible to businesses and end users.

Naturally, LLMs also greatly affect software development by offering rich capabilities to generate code from natural language descriptions and specifications. While general-purpose LLMs are already quite capable at this, there are also dedicated models~\cite{chen2021codex} for supporting software developers in their work. Products like GitHub Copilot\footnote{\url{https://github.com/features/copilot}}, Amazon CodeWhisperer\footnote{\url{https://aws.amazon.com/codewhisperer/}}, and many more are now being used by developers worldwide to enhance their productivity and improve the code and with that the software they write~\cite{Peng2023}.

One common way to integrate these models into the workflow is to prompt them in line in the editor as a richer, more context-sensitive auto-completion feature. While this has proven to increase developer productivity~\cite{Barke2022, Ross2023}, it is tailored towards quickly producing more code~\cite{Barke2022}. This interaction paradigm has not well supported other uses of LLMs for programming. An alternative is using a conversational user interface where the interaction with the LLM happens via text chat. Considering how LLM assistants are often considered an alternative to pair programming~\cite{copilotpairprogramming1,copilotpairprogramming2,copilotpairprogramming3,copilotpairprogramming4}, it makes sense to choose an interaction modality that one would also use with a coworker for interacting with the LLM assistant. Additionally, we see computational notebooks being extended with a wide range of extensions, for instance, graphical programming \cite{weber2024extending}. Thus, while this allows developers to interact more broadly with the LLM, the question of how best to integrate these interfaces into development tools like computational notebooks is still open.

In this work, we explore how conversational interfaces can extend and improve computational notebooks as one of the popular software development tools based on some existing approaches and feedback from practitioners and experts. We present a series of different design approaches that integrate conversational LLM support into computational notebooks to different degrees. We provide examples across a spectrum of interaction patterns and will discuss these interface designs' potential benefits and drawbacks.

\section{Extendings Computational Notebooks with LLMs}

In this section, we will briefly outline the far ends of the spectrum with standard computational notebooks without LLM support on the one side and pure conversational systems on the other. We will then explore various steps between these two poles and discuss their benefits and drawbacks. These concepts are based on both personal experience and unstructured discussions with practitioners and experts over the span of the last two years.

Computational notebooks like Jupyter notebooks\footnote{\url{https://jupyter.org}} or Google Colab\footnote{\url{https://colab.research.google.com}} follow the principles of literate programming and are built around the idea of mixing code with contextual and documentary ~\cite{KNUTH1984LITERATE, pieterse04, head2019, CN.Liu.2022, rule18, 10.1145/3377816.3381724, 9387490}. To this end, a software project is presented as a sequence of individual cells that can be either executable source code, the output of this code, arbitrary text, or even figures and visualizations~\cite{rule18, CN.Liu.2022}. While the interface is relatively simple, there is support for some mechanisms commonly found in other software developer tools, such as syntactical suggestions and auto-completion. As mentioned before, the code in the cells can be executed on demand. In this case, the code is typically sent to a backend of so-called ``kernels'' which execute the code~\cite{CLARKE2021100213, mage, watson2019pysnippet}, maintain the state of the program, etc. For the popular notebook-style environments, there are kernels for many programming languages. This makes the architecture of computational notebooks quite adaptable to many use cases. The interface itself can also be extended by adding new types of cells.

Still, at its core, computational notebooks closely resemble the old paradigm of the REPL (Read-Eval-Print-Loop), where the developer writes code, executes it, and then iterates upon their code. This requires intimate knowledge of the programming language in question and, in case of errors, a good understanding of the working principles of software code.

On the other hand, recent advancements in LLMs have made it possible for less experienced developers and complete novices to also quickly create software code and iterate on it when it does not perform as desired~\cite{Pearce2021, Sobania2023, Prenner2021, Trudova2020, Tufano2020}. This is possible because large language models can generate this output in response to simple language descriptions of the desired functionality. One prominent interface paradigm for this is the use of conversational systems. The user converses with the LLM in a chat style in these interfaces. To create a program, the user can request source code from the LLM via a prompt and, in subsequent prompts, refine it or request changes due to errors in other observations.

At first glance, these conversational systems and the chat histories they produce can look quite similar to computational notebooks: both feature an interleaving of natural language descriptions and code. They also both allow for rapid iteration of small code fragments. However, there are also some obvious differences between these two interfaces: since the computational notebook is built for software development, it comes with the whole infrastructure to quickly and easily execute the code via the kernels as described above. In conversational systems, this is commonly not the case, and the user has to choose an execution environment to see whether it works and does what it is meant to do.

The goal of the typical conversational UI is not to execute code but to give humans a means to interact with the machine in a natural format. While code cells in computational notebooks require a programming language with a rigid syntax, LLM chatbots can produce responses and potentially desired behavior even from vague, misspelled, or ambiguous descriptions in natural language.

So, both these interfaces strive to make two different aspects of software development easier and more accessible. Since notebook UIs offer an easy way to try out code and conversational UIs are an accessible way to create it, the obvious question is how to combine these two UI paradigms to maximize their benefits and find synergies.

\begin{figure*}[t]
    \centering
    \begin{subfigure}[b]{0.49\linewidth}
        \centering
        \includegraphics[width=1.0\linewidth]{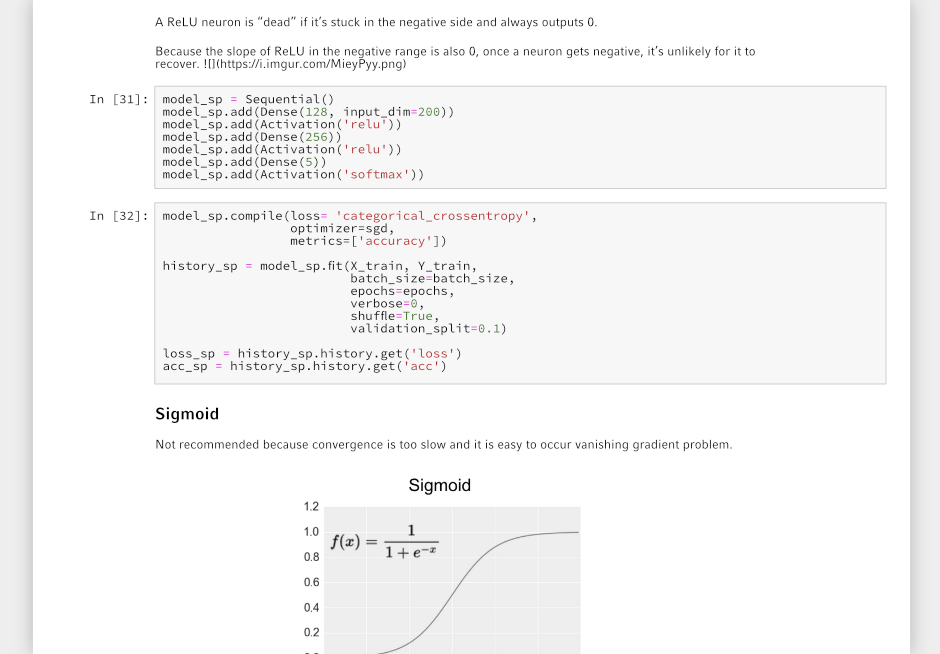}
        \caption{Without LLM support.}
        \label{fig:mockups-1}
    \end{subfigure}
    \hfill
    \begin{subfigure}[b]{0.49\linewidth}
        \centering
        \includegraphics[width=1.0\linewidth]{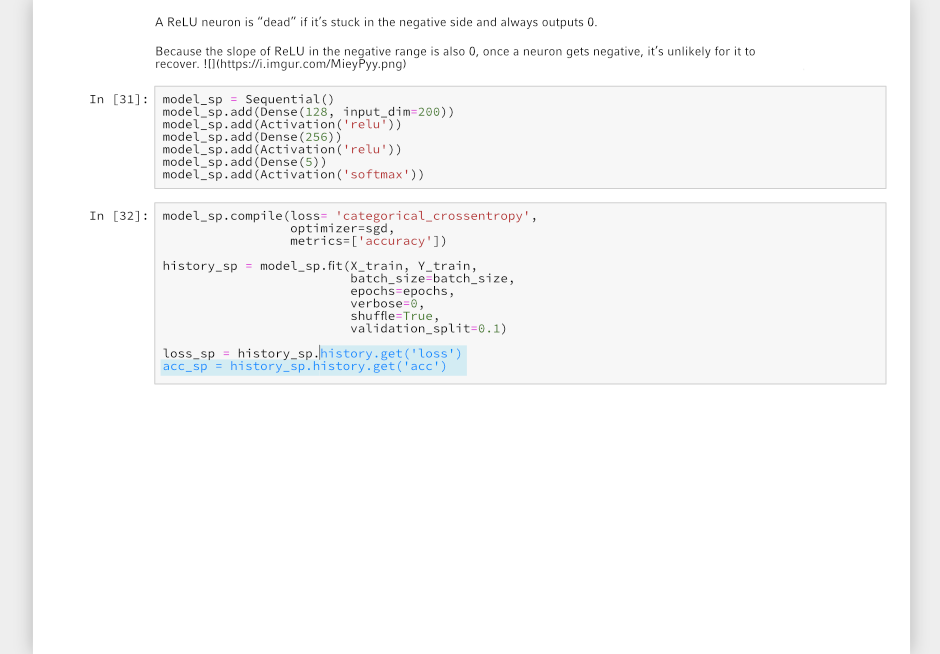}
        \caption{With inline LLM powered code completion.}
        \label{fig:mockups-2}
    \end{subfigure}

    \begin{subfigure}[b]{0.49\linewidth}
        \centering
        \includegraphics[width=1.0\linewidth]{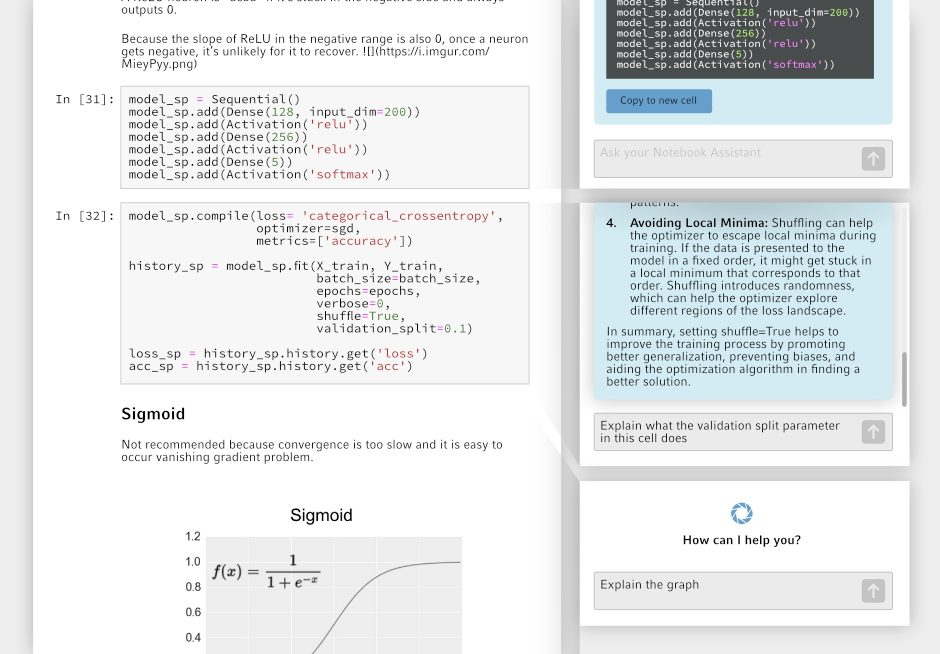}
        \caption{With per-cell LLM conversational support}
        \label{fig:mockups-4}
    \end{subfigure}
    \hfill
    \begin{subfigure}[b]{0.49\linewidth}
        \centering
        \includegraphics[width=1.0\linewidth]{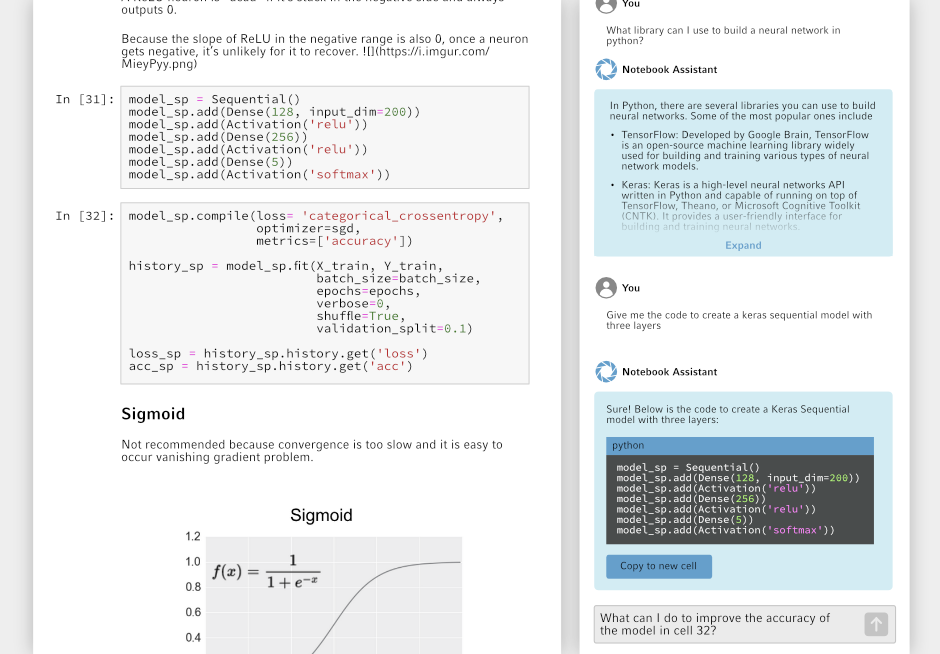}
        \caption{With global LLM conversational support}
        \label{fig:mockups-3}
    \end{subfigure}

    \begin{subfigure}[b]{0.49\linewidth}
        \centering
        \includegraphics[width=1.0\linewidth]{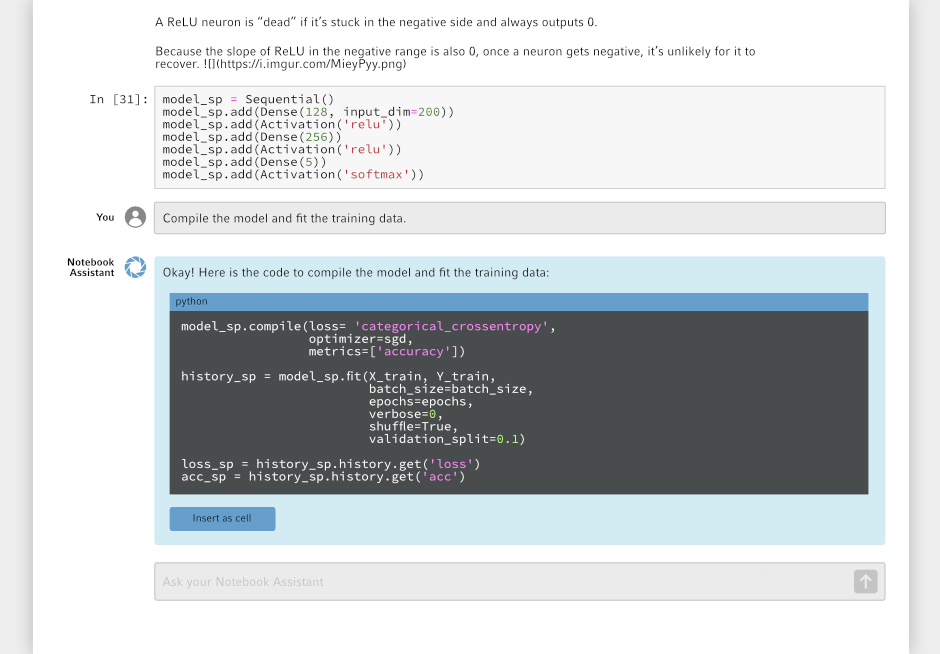}
        \caption{With inline LLM conversational support}
        \label{fig:mockups-5}
    \end{subfigure}
    \hfill
    \begin{subfigure}[b]{0.49\linewidth}
        \centering
        \includegraphics[width=1.0\linewidth]{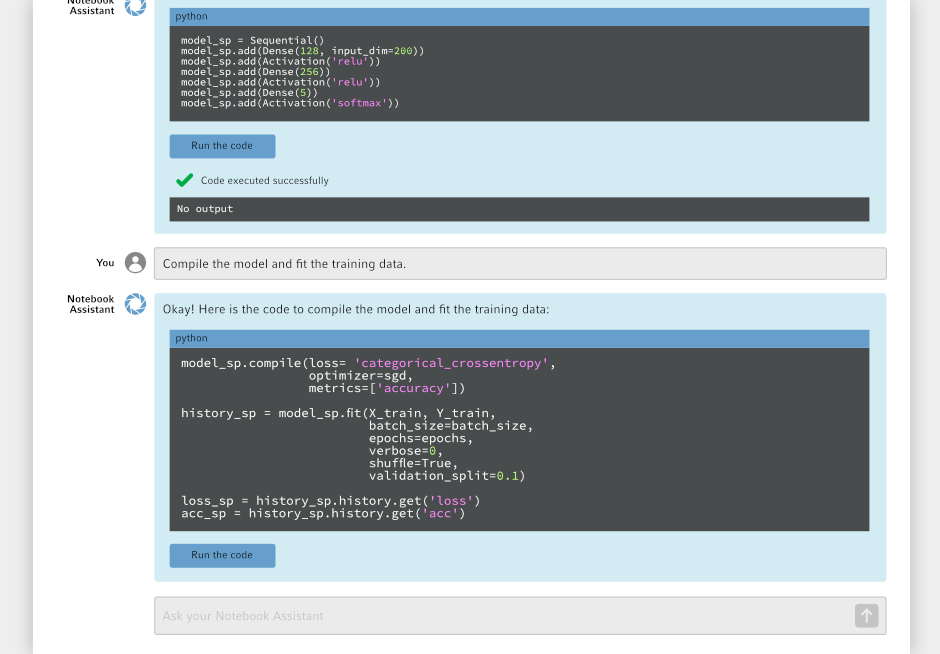}
        \caption{With inline LLM support and executable code responses.}
        \label{fig:mockups-6}
    \end{subfigure}

    \caption{The interfaces with increasing support of LLMs. a) shows the state of the art without LLM support, while b)-f) show the LLM support, which is highlighted in blue for illustration purposes.}
    \label{fig:mockups}
\end{figure*}

\paragraph{Executing LLM Responses}

Utilizing the infrastructure behind computational notebooks clearly is an easy way to implement this capability of executing code directly from within a chat. Like cells, the LLM code responses could be sent to a kernel backend, and the result becomes part of the chat history.
It may be noted that some vendors of conversational LLM interfaces offer the capability to execute the code directly from within the conversation\cite{lu2023chatgpt}, but this is usually still quite limited.
However, this is only one way to combine computational notebooks with the capabilities of LLMs into conversational notebooks (\autoref{fig:mockups-6}).

\paragraph{In-Cell Coding Assistance}

A second, fairly simple way, and one that is already being practiced, is to use the LLM-powered inline auto-completion of systems like GitHub Copilot to generate code and documentation in individual cells (\autoref{fig:mockups-2}). Anecdotal evidence suggests, however, that current implementations still struggle with considering the correct and complete context across cells, so this remains an area for future improvements.

Beyond this, a more comprehensive integration of these two paradigms is also possible. One next step towards a more conversational interaction style could be integrating the chat interface directly into the development environment, i.e., the computational notebook. This can be done side by side with the sequence of notebook cells (\autoref{fig:mockups-4}). Particularly when the chat interface is only used occasionally, requesting it on demand for specific cells seems a way to minimize distractions and visual clutter. This way, users can also have the chat history about a specific piece of code directly available without having to scroll through many unrelated messages to a specific cell.

\paragraph{Side-by-Side Interfaces}

However, suppose the chat considers the whole notebook as context. In that case, it is equally plausible to add a global chat next to the whole notebook and have these two interaction paradigms side by side (\autoref{fig:mockups-3}).
The Visual Studio Code extension for GitHub Copilot, for example, allows a layout that is very similar to that. However, so far, the chat is mostly a separate interface with only some functionality to simplify providing context to the LLM and copying back its results. The integration with the structure of notebooks is also still lacking at times, leading to limited options for interaction between the conversational agent and the notebook. Consequently, copying and pasting between the notebook and the chat is often necessary.

\paragraph{Conversational Notebooks}

To minimize the overhead and integrate conversational interaction more tightly into the notebook structure, we could add another cell type as a conversational cell (\autoref{fig:mockups-5}). Similar to how a developer would type code and the output is displayed below, in this cell type, the user could enter a prompt into the cell in the notebook, and the output below would be the response of the LLM.

\paragraph{Executable Conversations}

Since such a cell is then part of the program, we may treat it not just as a means to produce the code, but an ML-powered cell could directly alter the program state. In this case, we would skip the intermediate step of translating the natural language description to code and executing it. Instead, we would execute the natural language description directly. Effects and outputs of the execution could then be presented in natural language as well, leading to a seamless natural language conversation with computation in the background. In case of errors, there is still some value to debugging the code, but for the most part, it could be kept hidden, making notebooks more human-readable and accessible.\\

These different user interfaces all support the two key aspects, simplifying code creation and execution, and additional development tasks like documentation, explainability, collaboration, etc., to varying degrees. This emphasizes how there is no way to integrate LLM assistance into computational notebooks correctly. Instead, there appears to be a spectrum of how strongly the conversational aspects are integrated into computational notebooks or how the existing infrastructure of computational notebooks is utilized for the LLM (see~\autoref{fig:spectrum}). Depending on how this integration is executed in any novel conversational notebook interface, i.e., where it is placed on this spectrum, it will not just affect how to interact with it. Many aspects, like the ratio of natural language to programming languages, the granularity of the descriptions, etc., will give each interface distinct use cases and advantages and drawbacks.

\subsection{Advantages and Drawbacks}

While all these variants levels of conversational and LLM interaction in notebooks are viable, they will all be more suitable in some circumstances than others.

\paragraph{Inline Auto-Completion:}
Adding inline auto-completion promises a considerable productivity benefit for developers. Existing research~\cite{Barke2022} shows how particularly experienced developers can utilize this mechanism to build a piece of software using many small-scale completions quickly.
Since the code completions can be small, it is easier for the developer to maintain an overview of the actual generated code. Meanwhile, the time spent typing out large swaths of boilerplate code is greatly minimized. However, the speed increase may tempt some developers to breeze through the coding part and neglect documentation. While they may write comments to trigger auto-completion, these comments are intended for the LLM, so they are not necessarily phrased in a way that is most helpful for a human reader. Additionally, this interface is clearly tailored towards generating code. This interface is less ideal for ``discussing'' the model with the LLM, for example, asking for clarification or justification. However, prior studies demonstrated how developers can benefit from LLMs in this regard~\cite{Barke2022, Ross2023}. Conversational interfaces can also act as a partial replacement for extensively searching through API documentation and generally replace online search. Performing these tasks from within the code using code comments is sub-optimal at best.

\paragraph{Integrated Chat:}
These higher-level activities are where a separate chat interface alongside the computational notebook can shine, allowing for this kind of discussion in a dedicated space. As previously mentioned, both multiple smaller and larger global chats will add to the visual clutter and likely also to the mental workload of such an interface. Furthermore, the interaction between the notebook cells and the conversational interface is also an area where the interaction design will have a considerable impact. This interaction goes bi-directionally for sending contextual information from the notebook to the chat and integrating the responses into the notebook cells. However, the nature of a notebook can provide some valuable additional context for the LLM. For example, the model could automatically consider the distinction between code, output, and documentation cells to determine the users' intention, the approach, and the actual result. When an LLM considers this context, it makes sense to do this implicitly, not by copying redundant information into the prompt. However, this can make it less understandable, so mechanisms need to be in place to explain which context informed certain decisions in the code generation.

\paragraph{Challenges of the Notebook Paradigm:}
An additional open challenge remains the well-known fact that computational notebooks prescribe no strict execution order to the cells~\cite{head2019, lau2020thedesignspace}. This means that, e.g., through iteration, earlier cells are changed while later cells are outdated. This may lead an LLM to consider misleading context or suggest code changes that are detrimental to other code in the notebook. This further adds to the need for clear explainability of how the context played a role in the generated code.

When conversational cells are integrated into the notebook structure, one policy might be to consider just the previous cells as definitive context and assume that the following cells will be subject to change and are thus unreliable. However, the following cells will contain valuable context even if they become changed. Enforcing strict execution order of cells would theoretically solve this issue, but developers may consider this too restrictive and undesirable. However, this issue is one present for notebooks in general, not just conversational notebooks, so there should be broader research into adequate mechanisms to inform users and prevent or deal with errors due to execution order.

\paragraph{Accessibility vs. Accessing All Information}
Finally, treating natural language input as executable cells by using LLMs in the background can turn computational notebooks into mostly natural language documents. This makes it accessible for them to understand what a complete notebook does, but it will obscure the detailed inner workings. Particularly during development, this may be undesirable, and for debugging, it will be necessary to display the generated code. This way of integrating LLMs may be most viable if there is a mechanism in place to switch different views on a notebook, e.g., a development view that displays prompts, generated code, and outputs, including intermediate outputs, and a high-level view that displays the natural language description of the code and the relevant final outputs.
This kind of multi-view notebook might also support collaboration: first, a domain expert notes down the specification in natural language cells. Then, an LLM generates code from it, or a developer writes it with LLM support. After a debugging and refinement phase with the code, it is then translated back to natural language descriptions using LLM-based source code summarization, which can again be viewed by non-developers, e.g., domain experts, end users, etc.

\section{Future Work}

Evidently, combining computational notebooks with conversational interfaces has great potential to improve both interfaces. Both share the goal of making programming more accessible - conversational interfaces with LLMs that make it easier to produce code and computational notebooks by simplifying code execution and iteration. A combination of these interfaces can unlock synergies and create a user interface that offers the advantages of both while minimizing potential shortcomings.

Above, we outlined a number of interface concepts on the intersection of these UI types. We positioned these along a spectrum with varying degrees of conversational interaction and integration into notebooks. These only present a few ways to combine conversational interfaces with computational notebooks into conversational notebooks. We look forward to seeing how computational notebooks will improve in the future beyond some of the trivial ways that are currently being used, as listed above. In this endeavor, the spectrum we described should give authors of future conversational notebooks a point of reference to position their own interface concepts in relation to other design concepts.

Beyond designing and implementing more and different conversational notebooks, the obvious next step is to test whether these potential benefits actually hold true during real-world usage.
What metrics can determine whether conversational notebooks are a good interface will depend on a number of factors, though. Most important among those is the intended target group. As mentioned, both conversational interfaces and computational notebooks are means of making programming more accessible. Therefore, one possible target group is novices and domain experts who need to translate their expertise into software. For this target group, aspects of understandability, exportability, and overall positive user experience with little mental load seem desirable. Here, the goal might be a conversational notebook that lets domain experts focus on their area of expertise but guides them through the steps to produce the desired software. The ability to engage with the LLM in natural language to change existing code and ask for clarification or explanations will also help novice users to debug and iterate their software.

On the other hand, an improved development interface will also benefit professional software engineers. However, for this target group, an increase in productivity may be of greater interest. Of course, productivity can also include aspects that may benefit novices, like the ability to explain existing code. Beyond this, productivity for professional users also includes more quantifiable improvements like increased output, improved software quality, etc.~\cite{forsgren2021the}.

Consequently, it may very well be that novice and professional users will prefer the conversational aspect in their tools to different degrees.
Still, even for professional developers, it has been observed that they use LLM assistance in different modes~\cite{Barke2022}: when they rapidly request short, immediate suggestions, a well-integrated conversational system may be the top choice. When they are more in an ``explorative'' mode~\cite{Barke2022}, an open conversation may be the superior option, and a tight integration may be less important.

This points towards the fact that any exploration of the interface along the spectrum of conversational notebooks may have merits and maybe the top choice in certain scenarios and less ideal in others. Thus, it will be the task for researchers in this area to explore how to best utilize the synergies between computational notebooks and conversational systems and provide guidance to professional and amateur users alike when and how best to achieve their goals with conversational notebooks.

\balance
\bibliographystyle{ACM-Reference-Format}
\bibliography{main}

\end{document}